\documentclass[conference]{IEEEtran}
\IEEEoverridecommandlockouts
\usepackage{graphicx} 
\usepackage{amsmath,amssymb,amsfonts}
\usepackage{adjustbox}
\usepackage{amsmath,amssymb,amsfonts}
\usepackage{graphicx}
\usepackage{multirow}
\usepackage{cleveref}
\usepackage{algpseudocode}
\usepackage[caption=false,font=footnotesize]{subfig}

\usepackage{cite}

\usepackage{textcomp}
\usepackage{enumitem}
\usepackage{url}

\usepackage{underscore}

\usepackage{xcolor}

\title{Cluster-Weighted Training of Deep Surrogate Models for Subgrid Turbulent Transport }
\author{Rimsha Hameed Syeda, Dustin Kempton, Viacheslav Sadykov, \\ Irina Kitiashvili, Rafal Angryk}

\author{
\IEEEauthorblockN{Rimsha Hameed Syeda}
\IEEEauthorblockA{
\textit{Department of Computer Science} \\
\textit{Georgia State University} \\
Atlanta, Georgia \\
}
\and
\IEEEauthorblockN{Dustin Kempton}
\IEEEauthorblockA{
\textit{Department of Computer Science} \\
\textit{Georgia State University}\\
Atlanta, Georgia \\
}
\and
\IEEEauthorblockN{Viacheslav Sadykov}
\IEEEauthorblockA{
\textit{Department of Physics and Astronomy} \\
\textit{Georgia State University} \\
Atlanta, Georgia \\
}
\and
\IEEEauthorblockN{Irina Kitiashvili}
\IEEEauthorblockA{
\textit{Computational Physics Branch} \\
\textit{NASA Ames Research Center} \\
Moffett Field, California \\
}
\and
\IEEEauthorblockN{Rafal Angryk}
\IEEEauthorblockA{
\textit{Department of Computer Science} \\
\textit{Georgia State University}\\
Atlanta, Georgia \\
}
}

\date{August 2025}
\begin{document}
\maketitle
\begin{abstract}

Turbulence in the solar interior and atmosphere plays a crucial role in energy transport, yet modeling its subgrid-scale effects remains a major challenge. This study leverages machine learning (ML) models to predict components of the Reynolds stress tensor using high-resolution StellarBox simulations of the quiet Sun. Previously, we have compared a Multi-Layer Perceptron (MLP) and a 3D Convolutional Neural Network (CNN) against physics-based baselines to achieve a lower Mean Squared Error (MSE) and better generalization across various heights and depths in the solar atmosphere. To enhance learning, in this work, we investigate cluster-weighted training using K-Means and Hierarchical Agglomerative Clustering (HAC). By weighing the loss function based on cluster-specific prediction errors, we direct the model’s attention to high-error regions. It significantly improves CNN performance, achieving 34\% lower MSE and a significantly higher R² score indicating that integrating deterministic clustering with ML is a promising technique for modeling subgrid turbulence, in particular, and regression in diverse environments, in general.

\end{abstract}


\section{Introduction}\label{sec:intro}

Subgrid turbulence refers to the motions and interactions that occur at scales smaller than the grid resolution of a simulation and result in energy dissipation and momentum transport. Simulations typically maintain a trade-off between computational feasibility and ability to capture the high-resolution dynamics (both related to the characteristic scales of the modeled phenomena and the number of computational grid cells in each direction). In recent years, machine learning (ML) techniques, in particular, deep learning models, have shown promise in improving subgrid scale (SGS) models for magnetohydrodynamic (MHD) simulations by learning directly from high-fidelity high-resolution simulation data or observations \cite{Panda2021arXiv211107043P, Karpov2022}. Adapting these data-driven approaches can enable the development of more accurate and efficient surrogate SGS models for turbulent plasma and allows fast estimation of subgrid-scale turbulence feedback for a specific simulation run. It will significantly reduce the computational cost of MHD simulations for surrogate modeling purposes (such as exploration of the initial conditions space) without expensive additional computations involving high-resolution and more realistic physics-based turbulence models.

The overall idea of the current work follows the paper \cite{Karpov2022}, where high-resolution simulations of supernova explosions are utilized and the ability to run lower-resolution simulations by utilizing subgrid estimates of the higher-resolution properties is demonstrated. Their results show that the simulation domain behaves similarly to high-resolution simulations. In our follow-up work (\cite{syeda5236395developing}, briefly described in the current paper as well), we adopted their approach to simulations of the quiet Sun, utilizing realistic 3D solar radiative hydrodynamic simulations from \cite{Kitiashvili2015} obtained with StellarBox simulation code \cite{Wray2015_stellarbox, Wray2018vsss.book...39W}, along with exploring more sophisticated structures of deep learning models such as Multi-Layer Perceptron (MLP) and 3D Convolutional Neural Networks (CNN). Our results indicated that both MLP and CNN models can approximate the subgrid stress tensors more accurately with respect to the physics-based Gradient and Smagorinsky models. In the current work, we aim to further improve the accuracy of the CNN model by introducing a cluster-weighted training.

\section{Background}\label{sec:background}
In subgrid-scale turbulence modeling, the Reynolds stress tensor is a key quantity of interest, representing the correlation of the small-scale velocity field, which produces additional terms in MHD equations and can lead to kinetic energy dissipation and momentum transport. It is defined as:
\begin{gather}
    \tau{}_{ij} = \widetilde{u_{i}u_{j}} - \tilde{u_{i}}\tilde{u_{j}}
    \label{eqn:ReynoldsStress}
\end{gather}
Here $u_{i}$ is the $i$-th component of the velocity, $\widetilde{u_{i}u_{j}}$ represents the average of the product of two components, and~$\tilde{u_{i}}$ indicates the averaged value of the variable, both over the spatial scale of interest. We note here that very often the $u_{x}$ component of the velocity is noted in simulation approaches as $u$, the $u_{y}$ component as $v$, and the $u_{z}$ component as $w$. We will follow these notations in the current work.

Turbulence in the solar convection zone is anisotropic and inhomogeneous \cite{Nordlund1985} with convective scales increasing and flows becoming more homogeneous with depth, as shown in recent 3D radiative hydrodynamic modeling \cite{2025arXiv250200974K}. Traditional methods for estimating the Reynolds stress tensor in the convection zone rely on the spatial derivatives and statistics of the macroscopic flows, such as the gradient model \cite{Karpov2022}, the Smagorinsky model \cite{Smagorinsky1963_viscosity}, and the dynamic Smagorinsky model \cite{Germano1991_dynamicSmagorinsky}. As in \cite{syeda5236395developing}, here we utilize two physics-based subgrid models as the baselines for this study: the gradient model and the Smagorinsky model.

Recent studies such as \cite{ling2016reynolds,raissi2019physics} have explored deep learning for modeling turbulent flows. However, they typically use uniform training strategies that overlook spatial variability in model error. In contrast, our framework incorporates cluster-weighted training to explicitly account for heterogeneous regions in the input space.

\section{Dataset}\label{sec:data_prep}

\begin{figure*}[!t]
    \centering
    \includegraphics[width=0.95\linewidth]{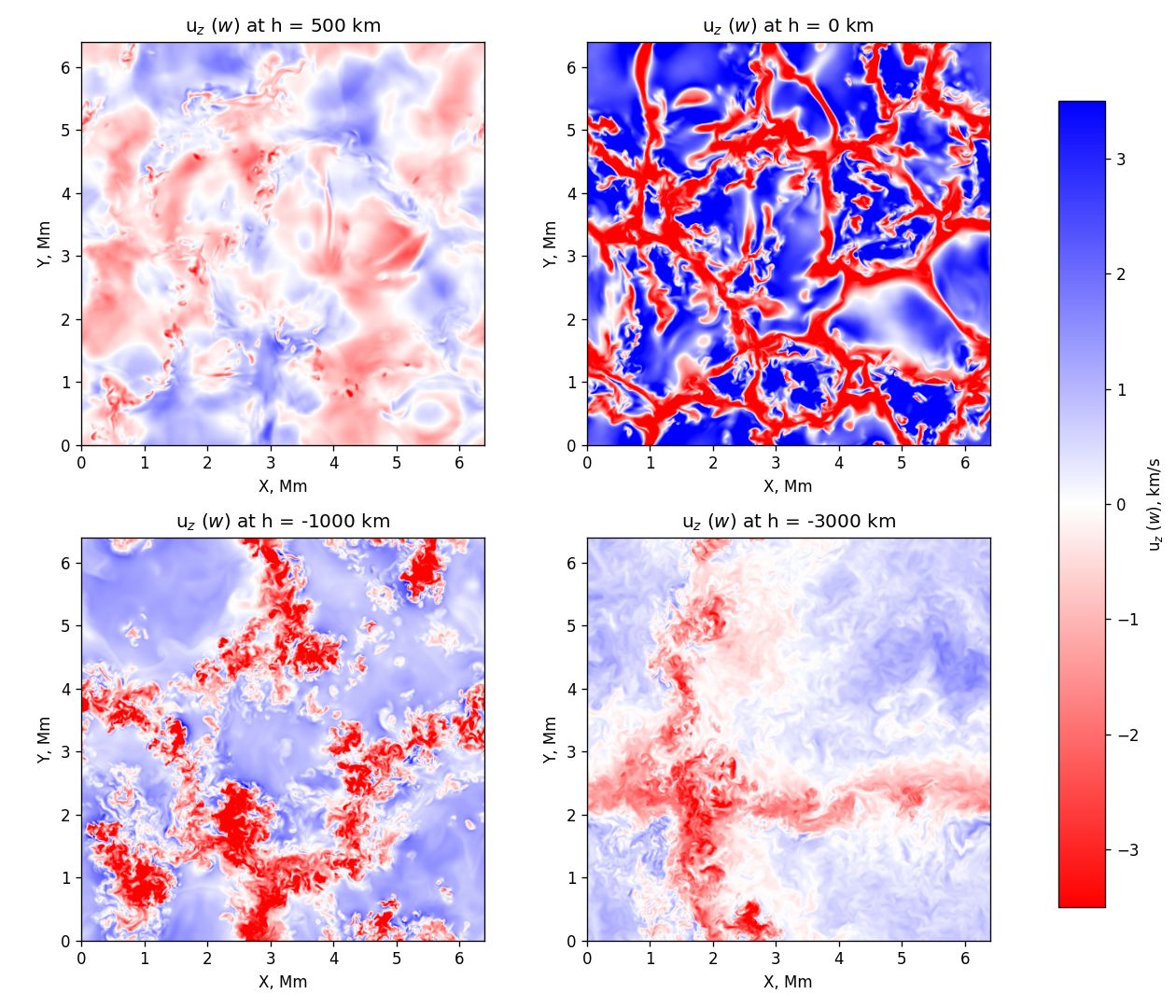}
    \caption{Illustration of vertical velocities in the simulations ($u_{z}$, or $w$) at the different heights of the simulation domain. One can note the changes in the spatial structures and velocity magnitudes when encountering different depths.}
    \label{fig:simulationvis}
\end{figure*}

We considered the 3D radiative hydrodynamic simulations with the StellarBox code that were previously used in \cite{Kitiashvili2015,Waidele2023ApJ...949...99W}. The simulated data are in the format of 512$\times$512$\times$512 size data cubes, with a spatial resolution of 12.5\, km in the horizontal scale (and a comparable uniform vertical scale), with a cadence of 30\,s. The data cubes have about 1\,Mm of the atmosphere above the geometrical $h=0\,$ km height, and $\sim$5.4\,Mm of the subphotospheric region. Because of the high resolution, we use these simulations to compute the Reynolds stress tensors directly (i.e., we do not assume any additional contribution from the Reynolds stresses at the subgrid-scale of this high-resolution model). The illustration of the behavior of the vertical velocity at different heights and depths of the simulations is displayed in Figure~\ref{fig:simulationvis}, reflecting different spatial structures of the plasma motions.

In order to train the ML SGS model, we must first extract and label the data from the existing simulation. The characteristic time scale at the solar photosphere is $\sim 5-10$ minutes. Therefore, we used every 10th data cube (or a 5-minute cadence) in the data series, a total of 25 data cubes, which allows us to make each considered data cube less dependent on the others. We then slice each data cube into 4$\times$4$\times$4 sub-cubes, effectively decreasing the resolution 4 times. For each third sub-cube in each direction, we compute the Reynolds stress tensor components directly following Equation.~\ref{eqn:ReynoldsStress}. We also compute the average density in this sub-cube, and average velocities in this and its 26 neighboring sub-cubes. The resulting task is to predict the $\tau{}_{ij}$ components in the central sub-cube based on their density and surrounding velocity fields. Because of the non-overlapping spatial sampling and sufficiently large temporal sampling, we ensure that our dataset consists of non-correlated data points and, therefore, could be split into train-validation-test datasets randomly. The final data set has $\sim$1.8\,M individual data points sampling the upper solar convection zone and a lower photosphere. Each training sample was represented as an 82-dimensional feature vector, incorporating the 3D neighborhood ($3 \times 3 \times 3$ cube) of three velocity components \( u, v, w \) and a scalar average density in the central cell. The six $\tau{}_{ij}$ components are targets. The preprocessing strategy closely follows the methodology in our previous work \cite{syeda5236395developing}. All input features were standardized using Z-score normalization using StandardScaler \cite{scikit-learn}, ensuring zero mean and unit variance between velocity and density variables. This normalization step equalizes feature influence and aids model convergence. We also transform the target variables ($\tau{}_{ij}$ components) to address strong skewness and variance instability. The diagonal components were log-transformed to improve normality, while off-diagonal components underwent a signed logarithmic transformation to preserve directional information. Finally, all transformed targets were standardized and range-scaled. This approach has been shown to stabilize training and improve model performance, particularly in regression tasks involving turbulent flow variables.

\section{Methodology}\label{sec:method}
\subsection{CNN Architecture and Training}\label{sec:cnnarch}

To model the non-linear relationship between local macroscopic flows and the subgrid Reynolds stress tensor, we design a 3D Convolutional Neural Network (CNN) trained on high-resolution StellarBox data. The CNN architecture used in this work is identical to that introduced in our previous study \cite{syeda5236395developing}. The $3 \times 3 \times 3$ cube of velocity components (\( u, v, w \)) are treated as separate channels passed through 3D convolutional layers to extract spatial features, which are then flattened and concatenated with the scalar density before entering fully connected layers that predict the six components of the Reynolds stress tensor. CNN was trained using Adam Optimizer with a learning rate of 0.001, a batch size of 128, with early stopping based on validation loss. Each convolution layer uses a $3 \times 3 \times 3$ kernel, and filter counts were increased in deeper layers to improve pattern recognition. The model was trained using a Mean Squared Error (MSE) loss function. For complete architectural and training details, refer to the Model Architecture and Parameter Tuning sections in \cite{syeda5236395developing}. The performance of the models was compared to both the Gradient and the Smagorinsky baselines across all stress components.

\subsection{Cluster-wise Error Analysis}\label{sec:error_analysis}

To investigate spatial variations in model performance and the feasibility of the initial motivation for the cluster-based approach described further, we perform a cluster-wise analysis on the CNN model trained with a standard uniform MSE loss. After training the baseline CNN following \cite{syeda5236395developing}, we apply K-Means clustering ($K=5$, Section~\ref{sec:inertia}) to the feature space (macroscopic velocities and the density) of the training samples. We then compute the Mean Squared Error (MSE) for each cluster based on the CNN’s predictions.

\begin{figure*}[!t]
  \centering
  \subfloat[]{\label{fig:clusteredloss}\includegraphics[width=0.5\textwidth]{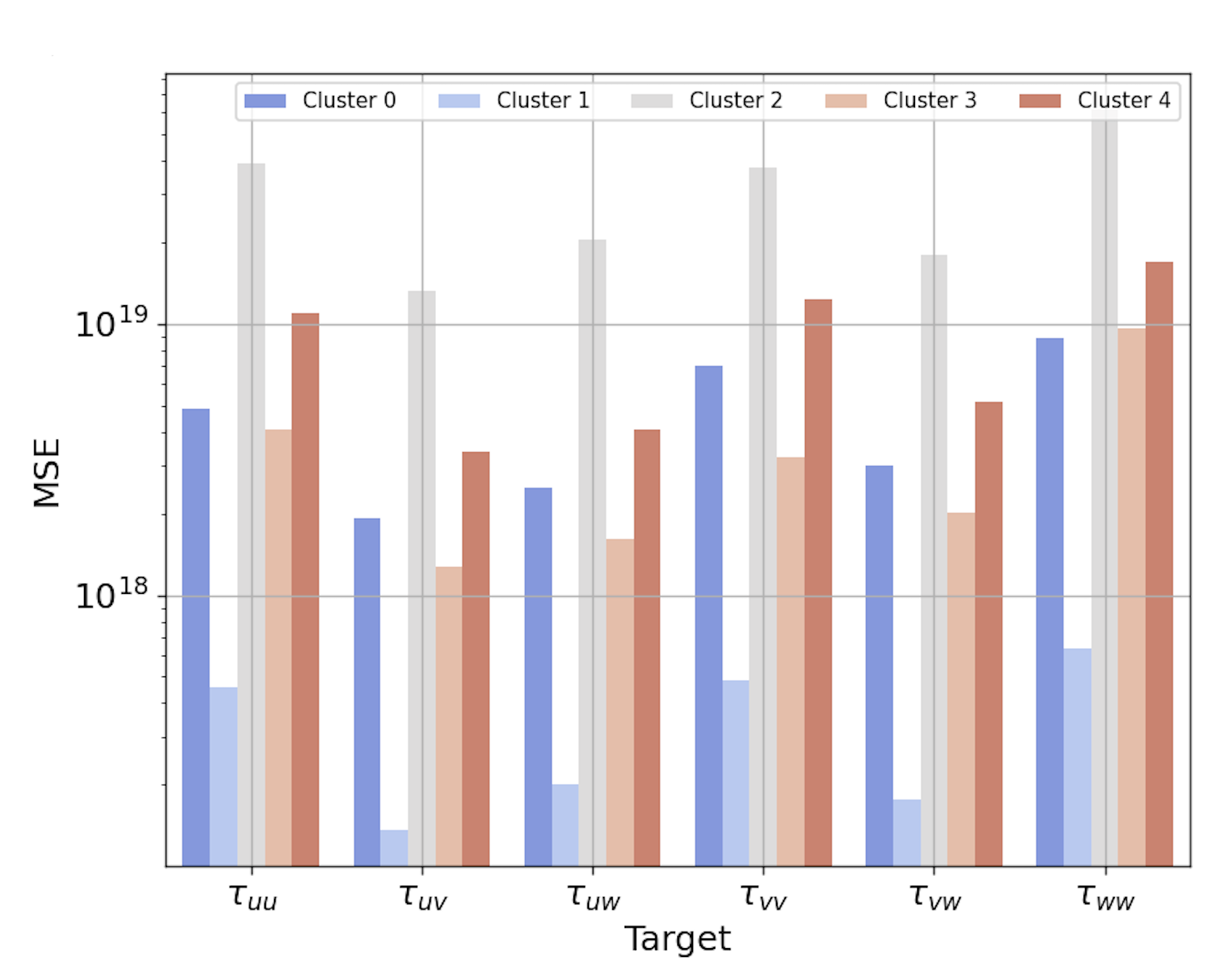}}\hfill
  \subfloat[]{\label{fig:clusterdist}\includegraphics[width=0.5\textwidth]{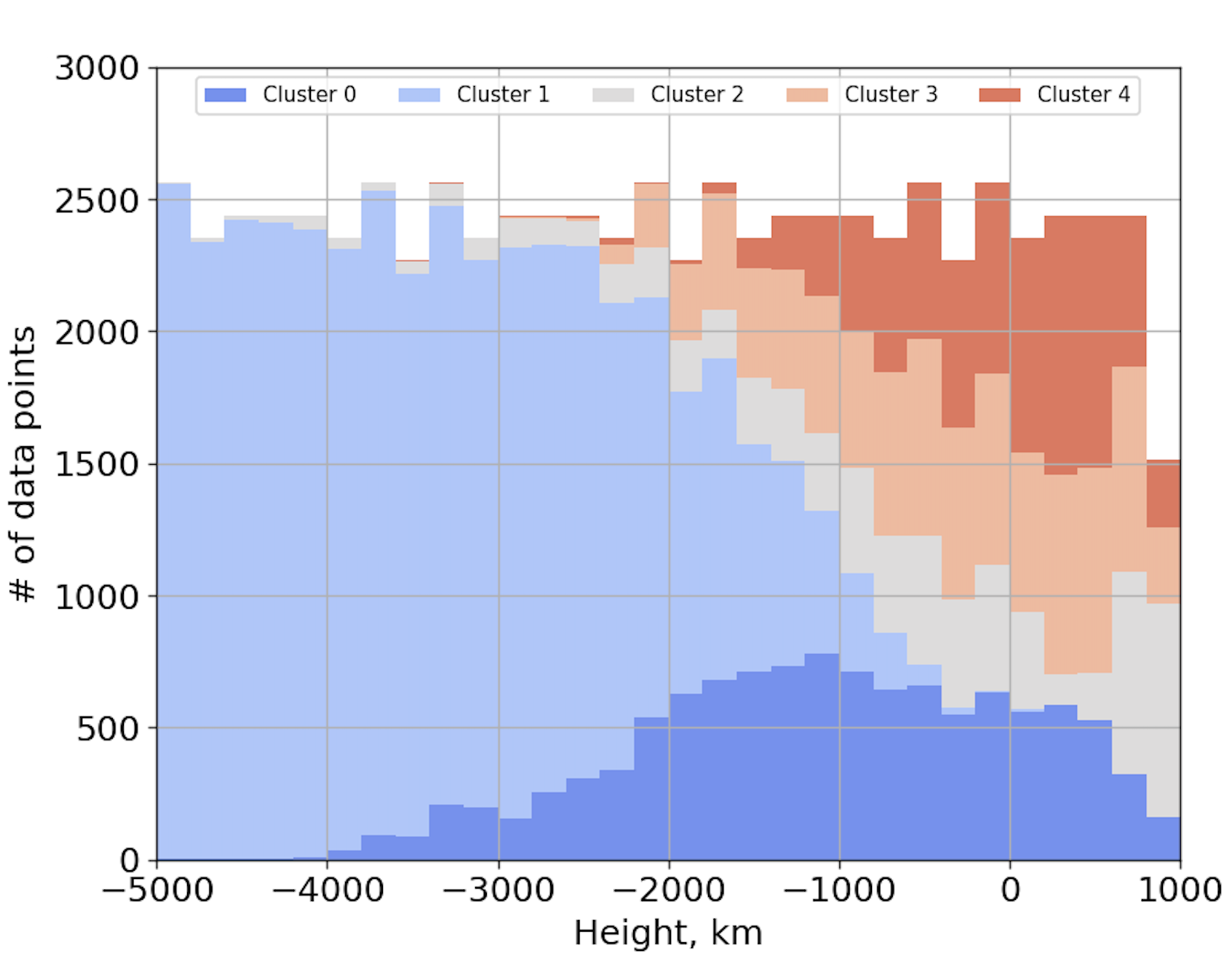}}
  \caption{\protect\subref{fig:clusteredloss} Cluster-wise MSE (cm$^{4}\!/\,\mathrm{s}^{4}$) for each $\tau_{ij}$ component using baseline CNN predictions.
  \protect\subref{fig:clusterdist} Distribution of clusters as a function of height.}
  \label{fig:cluster_mse}
\end{figure*}

Figure~\ref{fig:clusteredloss} presents the per-cluster MSE for each of the six $\tau_{ij}$ components, computed using predictions from the CNN-trained baseline. Although some clusters maintain relatively low error across all components, others consistently show elevated MSE. This indicates that certain regions of the input space are inherently harder to model, potentially due to more complex flow patterns or sharp gradients in velocity and density. The disparity in error magnitudes between clusters highlights the limitation of uniform loss, where all samples contribute equally to the loss function. One can also see from Figure~\ref{fig:clusterdist} that clustering does not isolate regions only based on their heights. While almost all points at the depth of $\sim$-5\, Mm belong to Cluster 0, the points in the solar atmosphere (here, 0-1\ Mm height) represent four different clusters. By identifying high-error clusters through this analysis, we justify the introduction of cluster-weighted training to prioritize learning in regions where the model underperforms.

\section{Implementation and Clustering Strategy}\label{sec:implementation}

\subsection{Clustering Method and Setup}\label{sec:setup}

To enhance regression accuracy using cluster-based loss weighting, we apply unsupervised clustering to partition the training data into representative subgroups. These clusters were then used to assign sample-specific weights in the loss function, with the aim of correcting imbalances in prediction error between inputs belonging to different clusters. Specifically, we aim to apply the K-Means algorithm to divide the training data according to their input characteristics (flattened or encoded), resulting in $k$ clusters ${C_1, C_2, \dots, C_k}$. This unsupervised segmentation allows us to assess prediction errors in structurally distinct input regions, facilitating targeted reweighting in subsequent training phases \cite{jain2010data}. Clustering is applied directly to the 82 feature vectors, described in Section ~\ref{sec:cnnarch}. The number of clusters was fixed at $k=5$ based on minimizing the average silhouette width scores, a metric that quantifies the quality of clustering by measuring cohesion and separation \cite{rousseeuw1987silhouettes}. Since K-Means clustering is sensitive to initial centroid selection, two variants of K-Means were evaluated to investigate the role of clustering quality and initialization strategies:
\begin{itemize}
    \item \textbf{Standard KMeans}: selects $k$ initial centroids uniformly at random from the data;
    \item \textbf{KMeans++}: selects centroid using probability proportional to its squared distance from the closest existing centroid \cite{Arthur2007kmeanspp}.
\end{itemize}
Standard K-Means can lead to poor local minima or unstable clustering. In contrast, K-Means++ improves this by choosing the initial centroids sequentially. This encourages diverse and well-separated initial seeds, often resulting in faster convergence and better final clustering outcomes. K-Means is inherently non-deterministic unless its random initialization state is fixed. For each method, we performed 20 runs with different random seeds to study the sensitivity of clustering outcomes to initialization.

\subsection{Inertia as Evaluation Criterion}\label{sec:inertia}
We used \textbf{silhouette scores} to determine the optimal number of clusters ($k$), selecting the value that maximized average silhouette width. However, to assess the stability and consistency of clustering across different random initializations, we used \textbf{Inertia}, the sum of squared distances from each sample to its assigned cluster centroid, as our primary diagnostic. Inertia is directly optimized by K-Means and provides a practical, interpretable metric for comparing runs in high-dimensional settings. This combination allowed us to ensure both meaningful cluster separation and reliable grouping behavior for use in our weighted training strategy.

\begin{figure}[!t]
    \centering
    \includegraphics[width=1.0\linewidth]{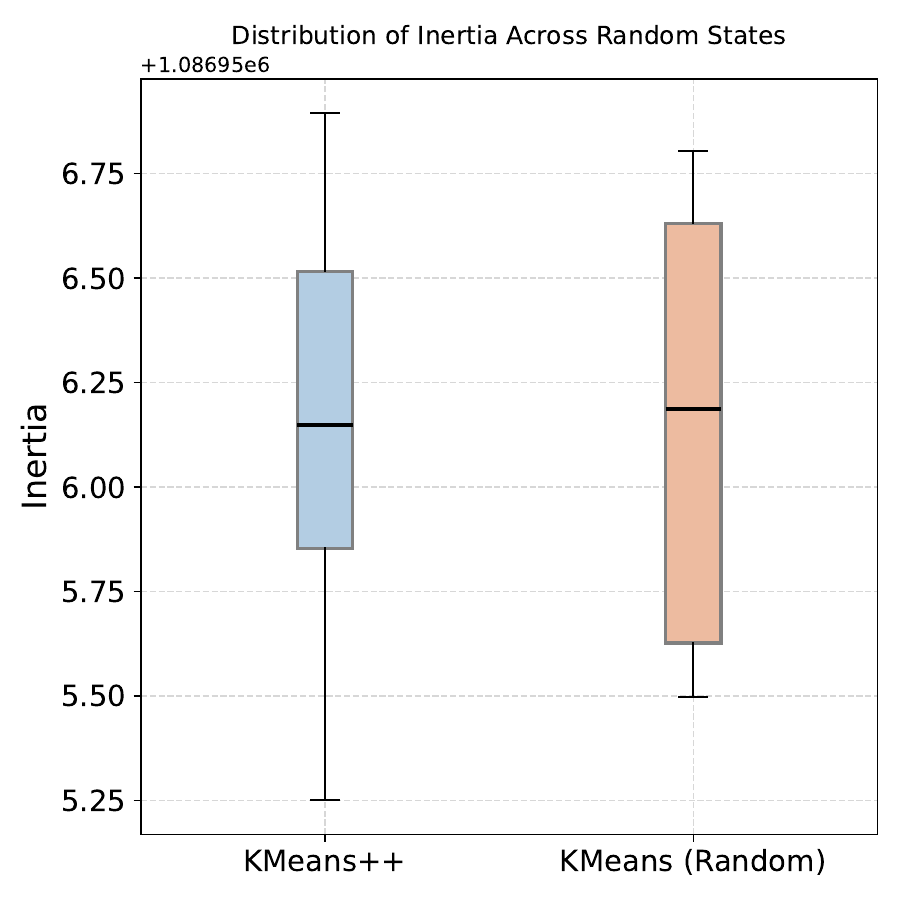}
    \caption{Distribution of inertia across random states using KMeans++ and random initialization.}
    \label{fig:inertia_comparison}
\end{figure}

Across all 20 runs, we found that the total inertia values were consistently centered around 1.08 million, with minimal variation across seeds. This indicated that our dataset exhibited stable clustering behavior, likely due to its large size (1.44 million samples used for clustering) and sufficient structure in the feature space. As expected, KMeans++ yielded slightly lower inertia on average compared to standard KMeans, demonstrating its ability to produce tighter clusters more reliably. As shown in Figure~\ref{fig:inertia_comparison}, both clustering methods produced highly stable results across initializations.

Given the small variance in inertia, we concluded that random initialization has little impact on the final cluster assignments in this application. However, K-Means++ was adopted for subsequent training due to its marginal but consistent improvements.

\subsection{Selection of Cluster Configurations for Training}\label{sec:cluster_configs}

From the 20 clustering runs, we selected five representative cluster configurations to use in training the cluster-weighted CNN:
\begin{enumerate}
    \item The clustering with the \textbf{lowest inertia}, representing the most compact clusters;
    \item The clustering with the \textbf{highest inertia}, capturing the worst-case compactness;
    \item Three additional clustering arrangements were sampled randomly from the remaining runs to ensure diversity.
\end{enumerate}

These configurations were then used to retrain the CNN model with a cluster-weighted loss function as defined in algorithm ~\ref{fig:cluster_weighted_training} in Section ~\ref{sec:weighted_train}. By comparing model performance across these clusterings, we were able to evaluate how the quality of clustering, as reflected by inertia, impacts the effectiveness of cluster-based training.

Despite selecting clustering configurations with both high and low inertia, we observed that the variation in model performance across these runs was relatively small. This reinforces the conclusion that the cluster-weighted training procedure is not highly sensitive to the specific initialization used in K-Means if the inertia variation is minimal. The consistently low variation in inertia across different seeds suggests that the input space has inherent structure that is reliably captured by the clustering algorithm, making the weighted loss framework robust and repeatable.

\subsection{Cluster-Weighted Training Method}\label{sec:weighted_train}

The weighting strategy shown in ~\ref{fig:cluster_weighted_training} is described below. After splitting the data into training and testing sets, we apply KMeans++ clustering ($K = 5$) to the normalized training set (velocities and density). CNN is first trained for $T_0=10$ epochs using standard MSE loss to obtain initial predictions. For each sample, we compute the total squared error across all targets, and calculate the mean error $MSE_k$ for the cluster $k$. Cluster weights $w_k$ are then assigned proportional to the relative error, scaled by a factor $\alpha = 3$, which emphasizes higher-error clusters without destabilizing training or overfitting the noisy samples. These weights are incorporated into a custom loss function, where each sample's contribution to the loss is multiplied by its cluster's weight.

\begin{figure}[!t]
  \centering
  \begin{minipage}{0.98\linewidth}\footnotesize
  \begin{algorithmic}[1]
    \Require Normalized input features $X$, target values $y$, number of clusters $K$, initial model $f_\theta$, scaling factor $\alpha$, initial training epochs $T_0$, maximum weighted training epochs $T$, early stopping patience $p$.
    \State Split $X, y$ into training and testing sets
    \State Apply KMeans++ on $X_{\text{train}}$ to obtain cluster labels $c_i \in \{1, \dots, K\}$
    \State \textbf{Initial Training:} Train $f_\theta$ for $T_0$ epochs using standard MSE loss:
    \[
    \mathcal{L}_{\text{MSE}} = \frac{1}{N \cdot T} \sum_{i=1}^{N} \sum_{j=1}^{T} \left( f_\theta(x_i)_j - y_{ij} \right)^2
    \]
    where $T$ is the number of targets per sample
    \State \textbf{Compute Cluster-Wise Errors:} For each sample $i$, compute the total squared error:
    \[
    \ell_i = \sum_{j=1}^{T} \left( f_\theta(x_i)_j - y_{ij} \right)^2
    \]
    \State Compute mean squared error $\text{MSE}_k$ for each cluster $k$:
    \[
    \text{MSE}_k = \frac{1}{|C_k|} \sum_{i \in C_k} \ell_i
    \]
    where $C_k$ is the set of samples in cluster $k$
    \State Compute weight for each cluster:
    \[
    w_k = 1 + \alpha \cdot \frac{\text{MSE}_k}{\max_j \text{MSE}_j}
    \] where $j$ indexes over $K$ clusters to compute max MSE
    \State Define custom weighted loss:
    \[
    \mathcal{L} = \frac{1}{N} \sum_{i=1}^N w_{c_i} \cdot \| f_\theta(x_i) - y_i \|^2
    \]
    \State Retrain $f_\theta$ using the weighted loss for $T$ epochs with early stopping p
    \State \Return Trained model $f_\theta$
  \end{algorithmic}
  \end{minipage}
  \caption{Cluster-Weighted Training}
  \label{fig:cluster_weighted_training}
\end{figure}

\section{Experiments \& Results}\label{sec:results}
\subsection{Alternative Clustering Strategies}\label{sec:hacdbscan}

In addition to K-Means, we explored Hierarchical Agglomerative Clustering (HAC) and DBSCAN to cluster the training data:

\begin{itemize}
    \item In \textbf{Hierarchical Agglomerative Clustering(HAC)}, clusters are formed by iteratively merging data points that minimize within-cluster variance. We experimented with HAC using Ward linkage, which iteratively merges pairs of clusters to minimize intra-cluster variance. This method is well-suited for uncovering hierarchical structure in the data, but is computationally expensive. Due to memory constraints, we were only able to perform cluster-weighted training using HAC on a subset comprising 2.5\% of the full training data. This limited the stability and resolution of the resulting clusters and prevented effective loss reweighting. As shown in Table~\ref{tab:hac_vs_kmeans}, even with power-based weighting ($w_c \propto \rho^{-p}$), the HAC-trained models significantly underperformed compared to both the KMeans-based model and the original CNN baseline. Despite this, the HAC experiments confirmed that error heterogeneity is a consistent feature of the training space, supporting our motivation for cluster-aware loss strategies.
    
    \item \textbf{DBSCAN} is a density-based clustering method that identifies arbitrarily shaped clusters and separates outliers. However, DBSCAN was computationally expensive, required sensitive parameter tuning (epsilon, minPts), and produced highly imbalanced clusters in high-dimensional feature space.
\end{itemize}

To directly evaluate performance, we implemented a variant of the cluster-weighted CNN using HAC-derived clusters and customized power-based weights proportional to inverse cluster density (e.g., $w_c \propto \rho^{-p}$). These results are summarized in Table~\ref{tab:hac_vs_kmeans}. While all methods confirmed the existence of the regions that were modeled with varying precision, K-Means offered more stable, balanced, and consistently improving clusters for loss reweighting. HAC-based weighting schemes underperformed even with tuning, and DBSCAN was discarded due to its instability in high-dimensional sparse data.

\begin{table*}[tb]
\caption{RMSE (cm$^{2}/$s$^{2}$) across $\tau_{ij}$ using CNN with uniform, K-Means, and HAC-based power-weighted loss schemes.}
\label{tab:hac_vs_kmeans}

\begin{center}
\begin{tabular}{lcccccc}
\textbf{Method} & \( \tau_{uu} \) & \( \tau_{uv} \) & \( \tau_{uw} \) & \( \tau_{vv} \) & \( \tau_{vw} \) & \( \tau_{ww} \) \\
CNN (uniform) & 2.36e+09 & 1.42e+09 & 1.67e+09 & 2.50e+09 & 1.70e+09 & 3.13e+09 \\
CNN (KMeans) & \textbf{1.65e+09} & \textbf{8.83e+08} & \textbf{9.93e+08} & \textbf{1.71e+09} & \textbf{9.92e+08} & \textbf{2.70e+09} \\
HAC, $p=2$ & 4.35e+09 & 1.69e+09 & 1.91e+09 & 4.43e+09 & 1.98e+09 & 5.28e+09 \\
HAC, $p=1$ & 3.99e+09 & 1.66e+09 & 1.95e+09 & 4.06e+09 & 2.01e+09 & 4.91e+09 \\
HAC, $p=0.5$ & 4.22e+09 & 1.71e+09 & 1.96e+09 & 4.09e+09 & 2.01e+09 & 4.31e+09 \\
HAC, $p=0.25$ & 4.34e+09 & 1.67e+09 & 1.96e+09 & 4.16e+09 & 1.99e+09 & 4.84e+09 \\
\end{tabular}
\end{center}
\end{table*}

\subsection{Comparison of PDF}\label{sec:pdf}

\begin{figure*}[!t]
  \centering
  \subfloat[\label{fig:pdftvv}]{
    \includegraphics[width=0.48\textwidth]{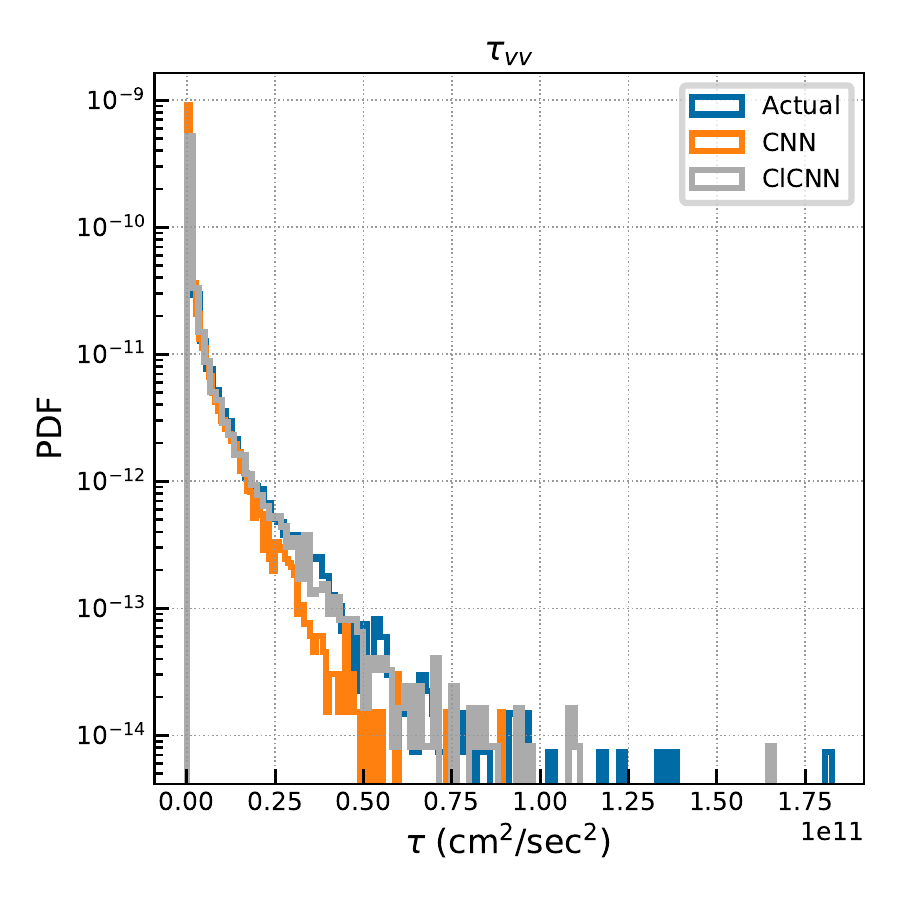}
  }\hfil
  \subfloat[\label{fig:pdftuw}]{
    \includegraphics[width=0.48\textwidth]{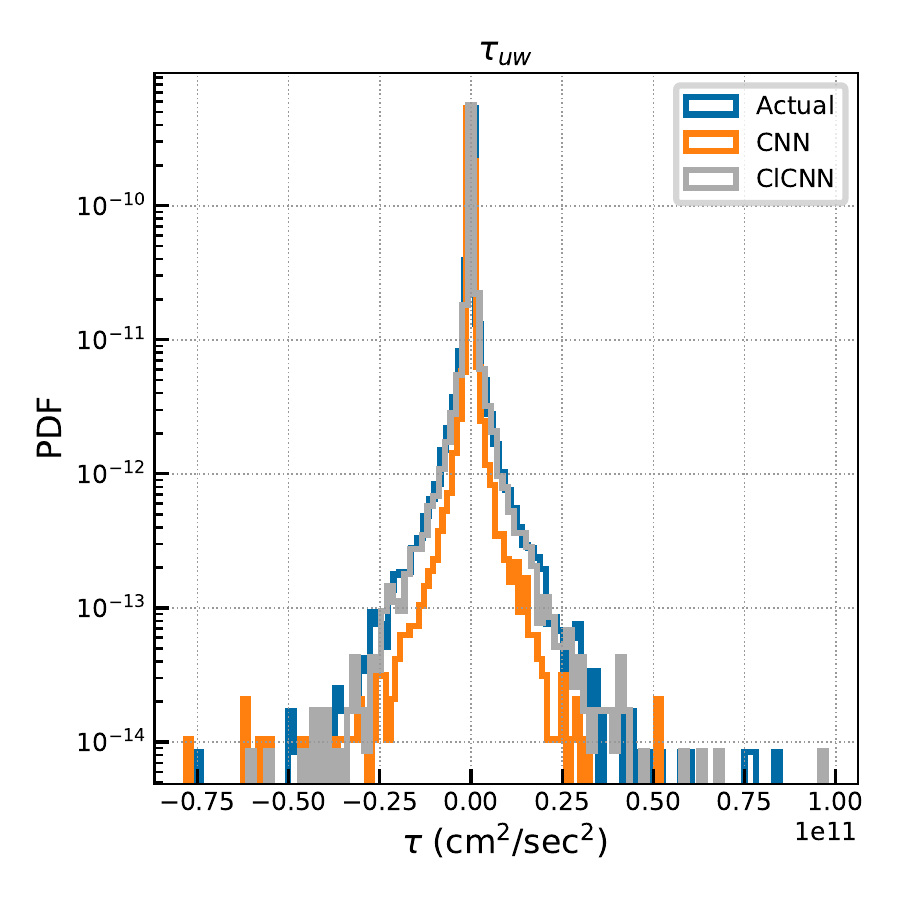}
  }
  \caption{Distributions of components $\tau_{vv}$ (a) and $\tau_{uw}$ (b).
  Blue curves show the original target data, orange shows the predictions of the CNN, and gray shows the predictions of the cluster-weighted CNN (ClCNN).}
  \label{fig:AllPDF}
\end{figure*}

The probability distribution functions (PDFs) of two different models (the baseline CNN and the cluster-weighted CNN, ClCNN) are visualized against the actual values of $\tau{}_{vv}$ and $\tau{}_{uw}$ Reynolds stresses in Figure  ~\ref{fig:AllPDF}. It is evident that the standard CNN underestimates extreme values and overconcentrates around the mean. In contrast, the cluster-weighted CNN more accurately captures both the peak and the heavy tails of the true distribution. This suggests improved modeling of rare or complex turbulent states — regions associated with higher training error and underrepresentation. These improvements are also pronounced in off-diagonal stress components (\( \tau_{uv}, \tau_{uw}, \tau_{vw} \), see Table~\ref{tab:hac_vs_kmeans}), which are more sensitive to local shear and density variations, reinforcing the effectiveness of cluster-based weighting in learning high-error regions. We note that in our previous work \cite{syeda5236395developing}, the improvement for modeling the off-diagonal components with respect to physics-based turbulence models was modest in comparison with the diagonal components.

\subsection{Average MSE and R\textsuperscript{2} Score Comparison}\label{sec:mse}
We compare the average Mean Squared Error (MSE) and R\textsuperscript{2} score across all six Reynolds stress components \( \tau_{ij} \) for each CNN model evaluated. The cluster-weighted CNN trained using KMeans clustering reduces average MSE by more than 34\% compared to the baseline CNN and shows improved generalization, in high-error regions.
\begin{table}[ht]
\centering
\begin{tabular}{p{4cm}cc}
\textbf{Model} & \textbf{Average MSE} & \textbf{Average R\textsuperscript{2}} \\
CNN (uniform loss) & $4.89 \times 10^{18}$ & 0.54 \\
CNN (cluster-weighted, KMeans) & $\mathbf{3.01 \times 10^{18}}$ & \textbf{0.80} \\
CNN (cluster-weighted, HAC on 2.5\% data) & $1.03 \times 10^{19}$ & 0.37 \\
\end{tabular}
\caption{Average MSE and R\textsuperscript{2} score across all $\tau$ components.}
\label{tab:avg_mse_r2}
\end{table}

\subsection{Error Distribution and Scatter Plot Comparisons}\label{sec:scatters}
\begin{figure*}[!t]
  \centering
  \subfloat[\label{fig:errortvv}]{
    \includegraphics[width=0.48\textwidth]{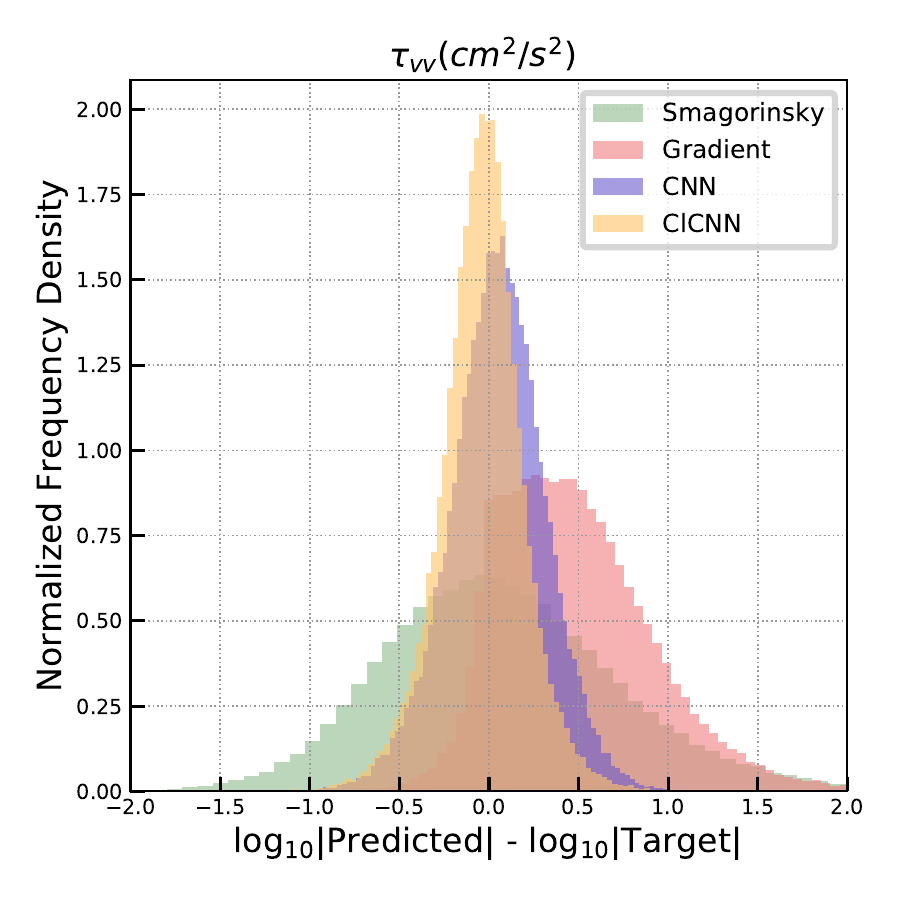}
  }
  \subfloat[\label{fig:errortuw}]{
    \includegraphics[width=0.48\textwidth]{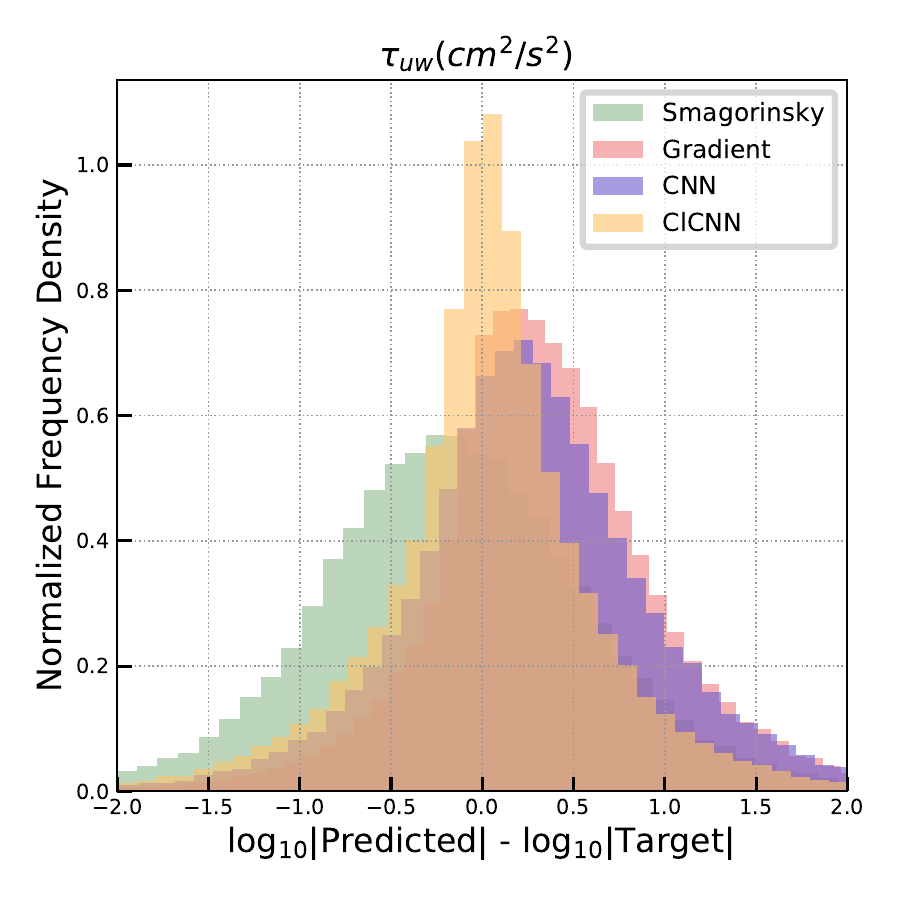}
  }
  \caption{Histogram of absolute prediction errors for $\tau_{vv}$ (a) and $\tau_{uw}$ (b).
  The cluster-weighted CNN (green) consistently reduces large error occurrences and tightens the error spread compared to the uniform-loss CNN (orange).}
  \label{fig:error_histograms}
\end{figure*}

We analyze the distribution of absolute prediction errors across models using histogram plots, and examine the scatter relationship between predicted and true values for key stress components. Figure~\ref{fig:error_histograms} compares the histograms of absolute errors for $\tau_{vv}$ and $\tau_{uw}$, highlighting differences in prediction accuracy across models. The standard CNN produces a wider error spread with a heavier right tail, especially for off-diagonal components like $\tau_{uw}$. The cluster-weighted CNN shifts the distribution leftward, reducing the frequency of large errors and improving concentration around zero. This indicates more stable predictions across turbulent regimes, particularly in underrepresented or high-variance regions.

\begin{figure*}[!t]
  \centering
  \subfloat[\label{fig:scattertvv}]{
    \includegraphics[width=0.48\textwidth]{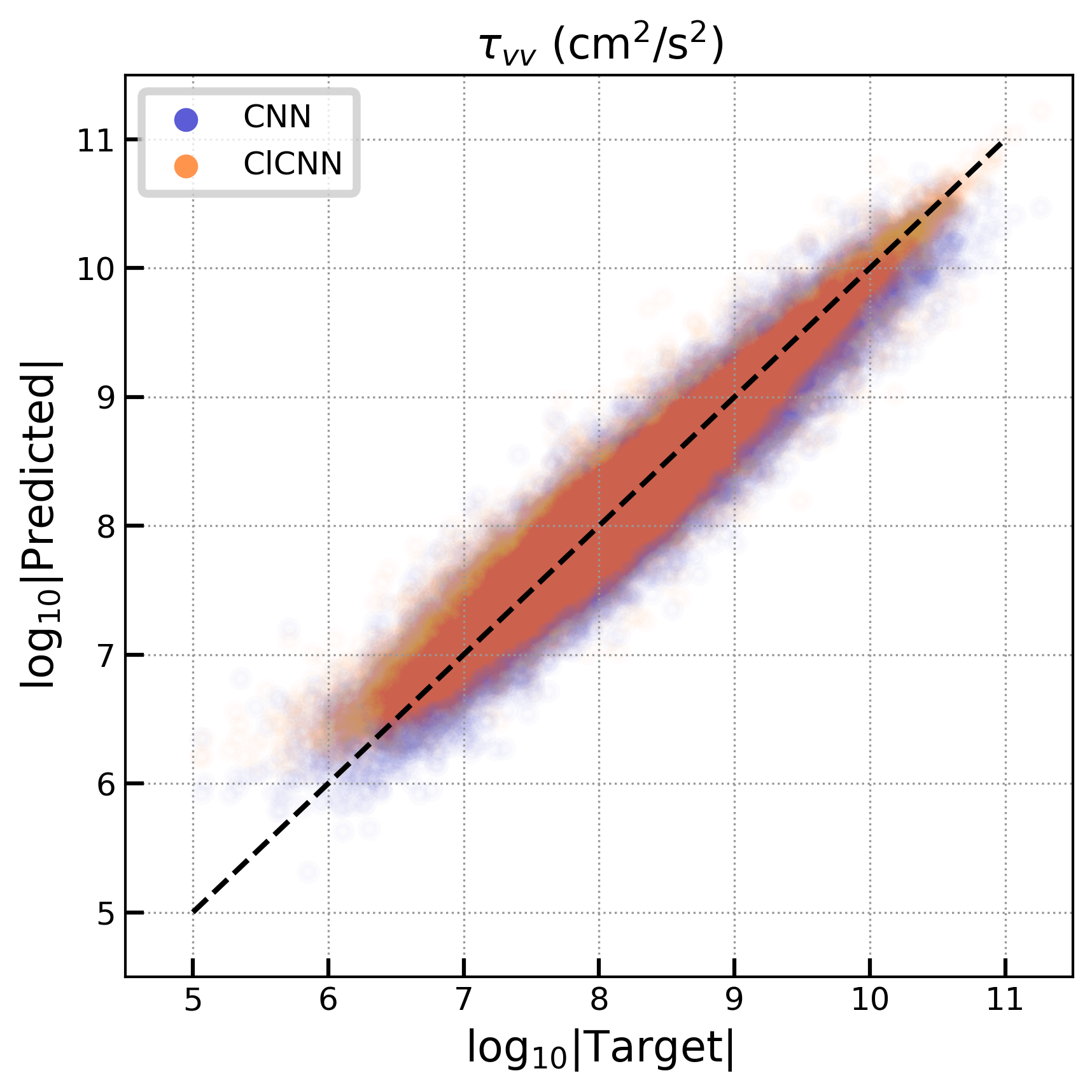}
  }\hfil
  \subfloat[\label{fig:scattertuw}]{
    \includegraphics[width=0.48\textwidth]{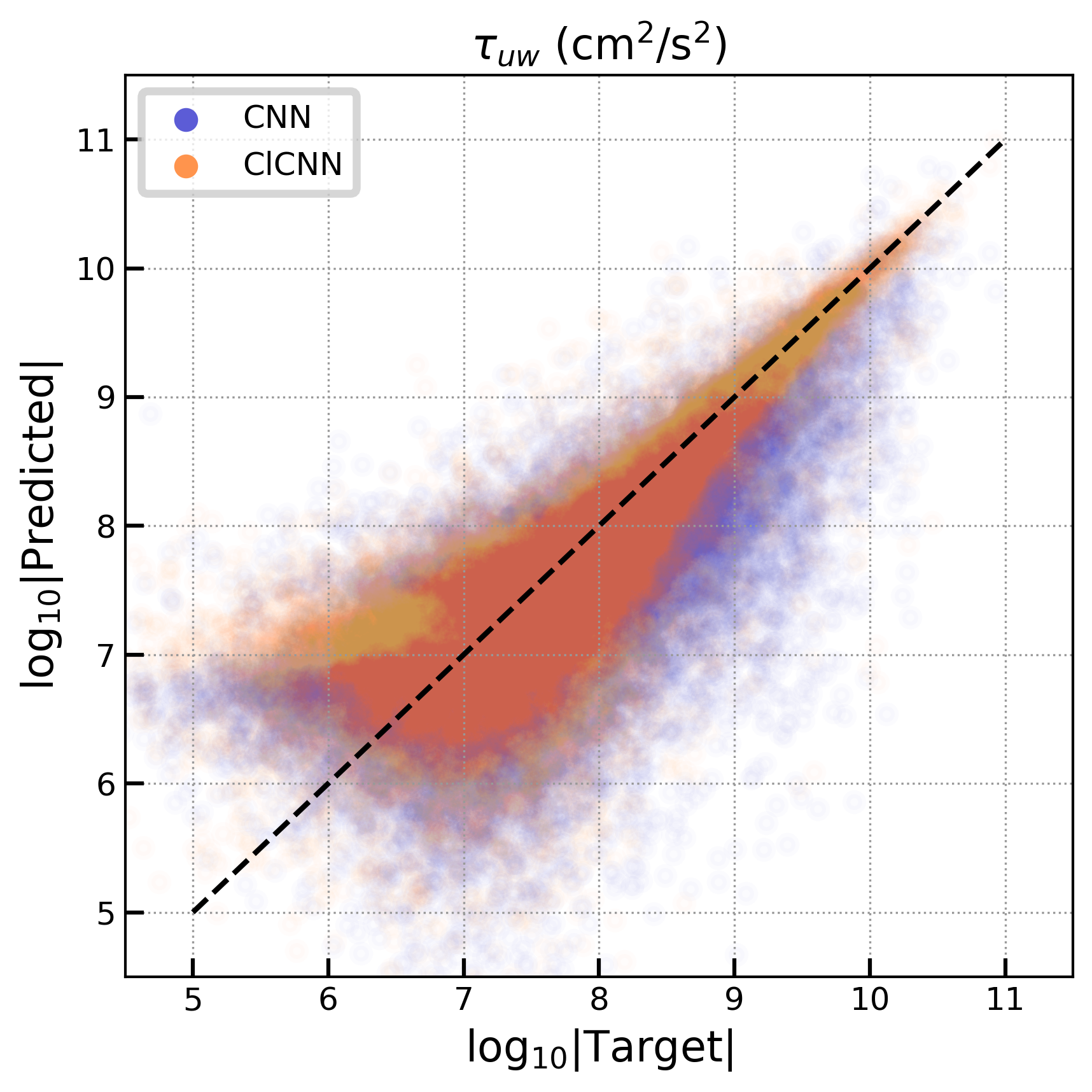}
  }
  \caption{Comparison of the predictions by the cluster-weighted CNN and the baseline CNN with the actual values of $\tau_{vv}$ (a) and $\tau_{uw}$ (b).}
  \label{fig:scatter_comparison}
\end{figure*}

To assess prediction fidelity, we plot predicted vs. true values for selected components in Figure~\ref{fig:scatter_comparison}. The CNN underpredicts large-magnitude values and exhibits vertical scatter, especially in low-density regimes. The cluster-weighted CNN(ClCNN) achieves tighter alignment with the identity line, indicating more consistent recovery of true stress magnitudes. Outliers are reduced, suggesting better handling of edge cases and improved robustness in complex flow regimes. These results further confirm that incorporating cluster-aware loss leads to more accurate and physically consistent predictions, particularly in regions of the data that were previously underfit.

\begin{figure*}[!t]
  \centering
  \subfloat[\label{fig:med_rel_mse}]{
    \includegraphics[width=0.48\textwidth]{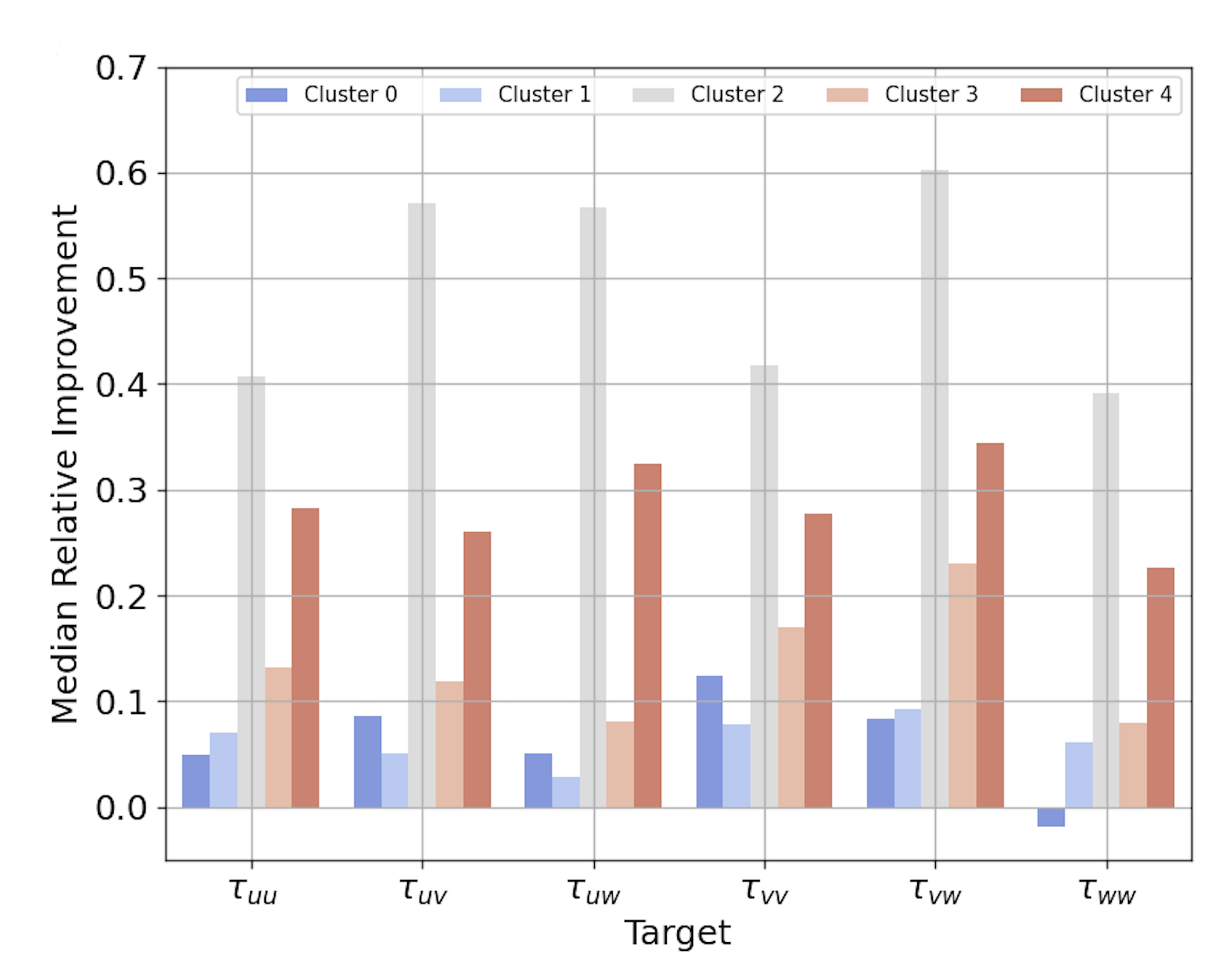}
  }\hfil
  \subfloat[\label{fig:med_rel_ht}]{
    \includegraphics[width=0.48\textwidth]{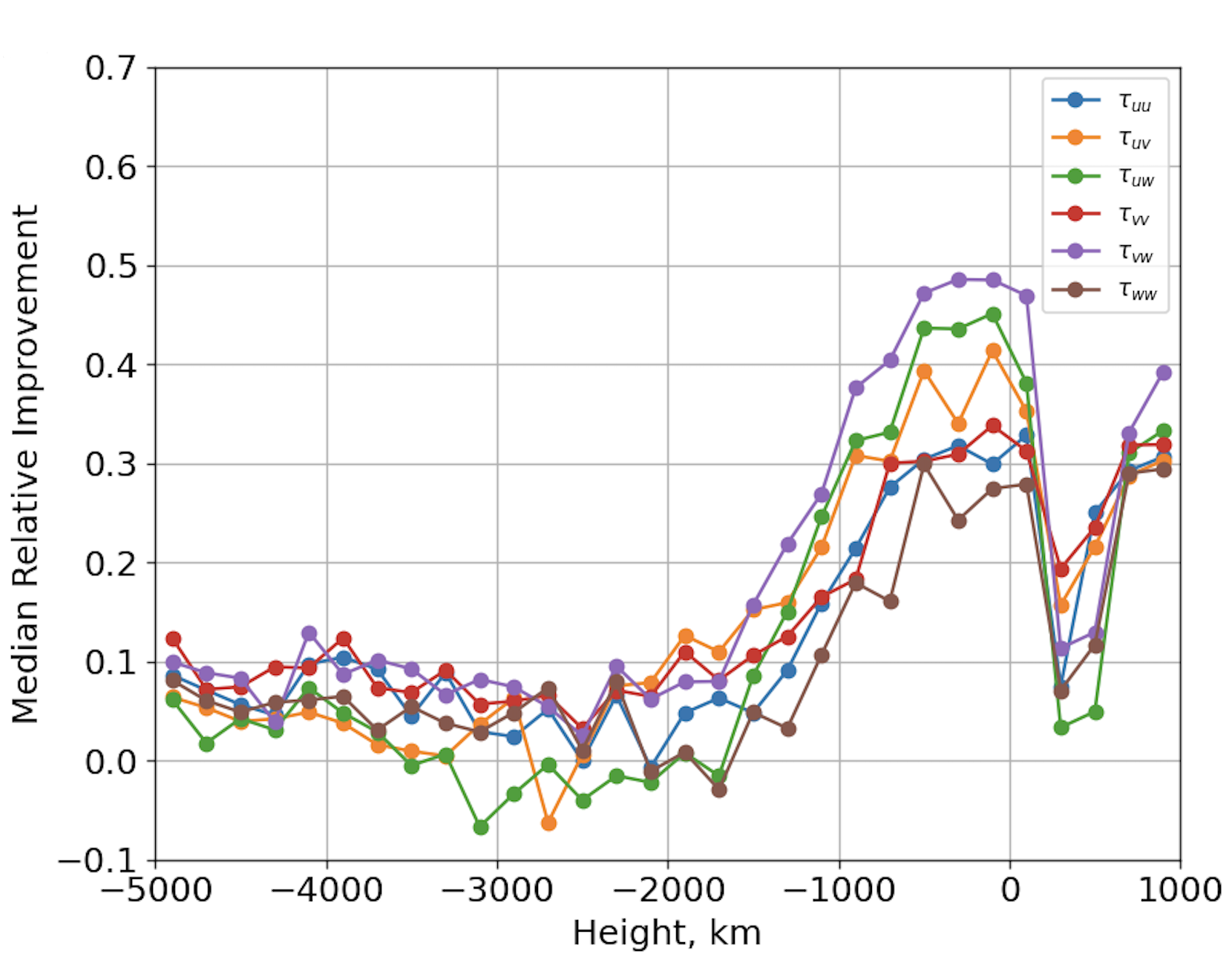}
  }
  \caption{Cluster-wise median relative improvement of MSE (a) and relative improvements as a function of height (b) across all $\tau$.}
  \label{fig:clusters_imp}
\end{figure*}

Figure~\ref {fig:clusters_imp} illustrates the median relative improvement of the cluster-weighted CNN over the baseline CNN model. The Figure~\ref{fig:med_rel_mse} shows improvements across clusters for each Reynolds Stress Tensor component, where certain clusters, 2 and 4, exhibit a reduction in median error of approximately 30-60\%. This confirms that the error heterogeneity is strongly linked to input data segmentation. The analysis in Figure~\ref{fig:med_rel_ht} localizes error reduction in physically meaningful height ranges with the most pronounced gains occurring between -1000 and 0 km, where errors are reduced up to 50\%. These regions correspond to the upper convection zone and lower atmosphere, which are typically harder to model due to sharp gradients and complex turbulent flows.

\section{Conclusion and Future Work}\label{sec:conclusion}
In this study, we introduced a cluster-weighted training framework to improve the accuracy of a deep learning surrogate model for subgrid turbulent transport. The model is trained on data derived from high-resolution 3D radiative hydrodynamic simulations of the quiet Sun. Building upon our previous CNN-based surrogate \cite{syeda5236395developing}, we improved model performance by incorporating a K-Means-driven loss reweighting scheme that targets regions with high prediction error. The identified regions were heavily weighted in the loss function during retraining. This strategy led to a significant improvement in model performance across all components of the Reynolds stress tensor, reducing the average MSE by over 34\% and improving the R\textsuperscript{2} score from 0.54 to 0.80 compared to the CNN baseline. The improvements were notable for off-diagonal components, which are more sensitive to local shear and density variations. We also compared alternative clustering techniques, including HAC and DBSCAN, which revealed error heterogeneity, but were less effective due to memory limitations and high-dimensional instability. K-Means was found to be the most robust and computationally efficient in our setting. This work demonstrates that targeted loss reweighting using unsupervised clustering can substantially improve surrogate model generalization in turbulent flow prediction. In practice, the weighted training added an overhead of 15–20\% due to clustering and reweighting, while the inference time remains identical to the baseline since the architecture is unchanged. The analysis also highlights that cluster-weighted training not only enhances global metrics but also yields significant localized benefits in both feature-space-defined clusters and physically relevant height ranges.

Future work will explore: (1) evaluating model generalization across varying spatial and temporal resolutions to assess scale sensitivity; (2) incorporating magnetic field features to enhance surrogate accuracy in magnetoconvection settings \cite{Panda2021arXiv211107043P, Rempel2018}; (3) investigating alternative architectures such as Vision Transformers (ViTs) to better capture long-range spatial dependencies in turbulent flows; and (4) comparing cluster-weighted training with uncertainty-based sampling strategies, aiming to develop an adaptive training framework that prioritizes informative, high-error samples. In the long term, our goal is to incorporate the developed models directly into the StellarBox simulation code in place of the current physics-based subgrid turbulence models. This will also allow us to perform the proper physical validation as well, by directly comparing the lower-resolution simulation runs with the ML surrogate subgrid turbulence models and the higher-resolution simulation runs initiated with the same boundary and initial conditions.

\section*{Acknowledgment}

This project has been supported in part by funding from NASA's COFFIES DRIVE Science Center grant 80NSSC22M0162. It has also been supported in part by the NSF Faculty Development in geoSpace Science program through grant 1936361.

\bibliographystyle{IEEEtran}
\IEEEtriggeratref{14}
\bibliography{refer} 
\end{document}